\input{aipcheck}

\documentclass[
    ,final            
  ]
  {aipproc}

\layoutstyle{6x9}

\usepackage{overpic}
\usepackage{subfigure}
\usepackage[utf8]{inputenc}
\usepackage[english]{babel}
\usepackage{amsmath}
\usepackage{amsfonts}
\usepackage{amssymb}
\usepackage{epsfig}
\usepackage{graphics,psfrag,rotating}
\usepackage{graphicx}
\usepackage{dcolumn}
\usepackage{bm}

\begin{document}

\title{Modified spinning black holes}


\author{J.\,A.\,R.\,Cembranos}{
  address={Departamento de F\'{\i}sica Te\'orica I, Universidad Complutense de Madrid, E-28040 Madrid, Spain.}
}

\author{A.\,de la Cruz-Dombriz}{
  address={Astrophysics, Cosmology and Gravity Centre (ACGC), University of Cape Town, Rondebosch, 7701, South Africa.}
 ,altaddress={ Department of Mathematics and Applied Mathematics, University of Cape Town, 7701 Rondebosch, Cape Town, South Africa.}
}

\author{P. Jimeno Romero}{
address={Departamento de F\'{\i}sica Te\'orica I, Universidad Complutense de Madrid, E-28040 Madrid, Spain.} 
}

\begin{abstract}
In the context of $f(R)$ modified gravity theories we determine that the black holes existence is determined by the sign of a parameter 
dependent of the mass, the charge, the spin and the scalar curvature. We obtain the different astrophysical objects derived from the presence of different horizons. Thermodynamics for this kind of black holes is studied for negative values of the curvature, revealing that we can distinguish between two kinds of BH: {\it fast} and {\it slow}.
\end{abstract}

\maketitle


\section{Black Holes Structure and Thermodynamics}

Black holes (BH) properties have been widely analyzed in many different gravitational theories \cite{cvetic,Mignemi,Hawking,Dombriz_varia}.
Although General Relativity (GR) has been the most successful gravitational theory, different questions, such as the necessary introduction of dark matter \cite{DM,WIMPs,isearches,Coll} and dark energy \cite{DE,cosmoproblema} to describe the late expansion of the Universe, suggest to study alternative gravitational interactions \cite{varios}, and in particular, $f(R)$ theories inside the metric formalism \cite{fR}.\\

Since we are looking for constant curvature $R_0$
vacuum solutions for fields generated by massive charged
objects, the appropriate $f(R)$ action leads to the modified field equations:
\begin{equation}
 R_{\mu\nu}\,(1+f'(R_0))-\frac{1}{2}\,g_{\mu\nu}\,(R_0+f(R_0))-2\left( F_{\mu\alpha}F_{\,\,\,\nu}^\alpha-\frac{1}{4}g_{\mu\nu}F_{\alpha\beta}F^{\alpha\beta}\right) \,=\,0\,\nonumber
\label{ec_tensorial}
\end{equation}

In Boyer-Lindquist coordinates, the axisymmetric, stationary and constant curvature $R_0$ solution of the BH with mass, electric charge and angular momentum, takes the form:
\begin{eqnarray}
\text{d}s^2\,=\,\frac{\rho^2}{\Delta_r}\,\text{d}r^2+\frac{\rho^2}{\Delta_\theta}\,\text{d}\theta^2+\frac{\Delta_\theta\,\sin^2{\theta}}{\rho^2}\,\left[a\,\frac{\text{d}t}{\Xi}-\left(r^2+a^2\right)\,\frac{\text{d}\phi}{\Xi}\right]^2-\frac{\Delta_r}{\rho^2}\,\left(\frac{\text{d}t}{\Xi}-a\sin^2{\theta}\frac{\text{d}\phi}{\Xi}\right)^2\,\nonumber
\label{metrica}
\end{eqnarray}
with: $\Delta_r\,:=\,\left(r^2+a^2\right)\left(1-{R_0}/{12}\,r^2\right)-2Mr+{q^2}/{\left( 1+f'(R_0 )\right)}$, $\rho^2\,:=\,r^2+a^2\cos^2{\theta}$, $\Delta_\theta\,:=\,1+{R_0}/{12}\,a^2\cos^2{\theta}$, $\Xi\,:=\,1+{R_0}/{12}\,a^2$, where $M$, $a$ and $q$ (with $Q\equiv q\, /\,\left(1+f'(R_0)\right)^{1/2}$) denote the mass, spin and electric charge parameters respectively, and the scalar curvature $R_0$ is given by $R_0=2\,f(R_0)/(f'(R_0)-1)$.

The horizons are found as the roots of the fourth order equation $\Delta_r=0$. The existence 
of real solutions for this equation is given by the so called {\it horizon parameter}: $h$ \cite{Cembranos:2011sr}. It is remarkable that, from a certain positive value of the curvature $R_{0}^{\,crit}$ onward, the $h$ factor goes to zero for two values of $a$, i.e., apart from the usual spin for which the BH turns $extremal$, there is now a spin lower bound, below which the BH turns into a {\it marginal extremal} BH.\\

\begin{figure}[tp]
	\centering
		\includegraphics[width=0.41\textwidth]{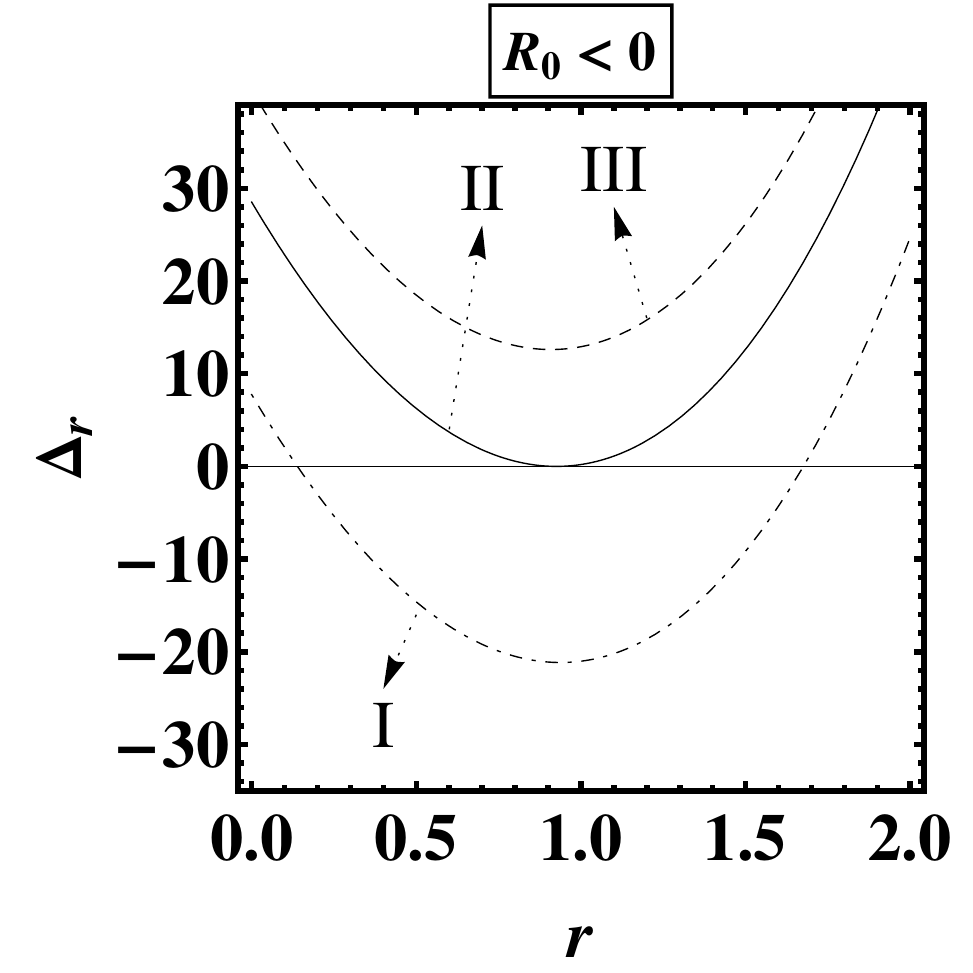}\,\,\,\,\,\,\,\,\,\,\,\,\,\,\,\,\,\,\,\,\,\,\,
		\includegraphics[width=0.41\textwidth]{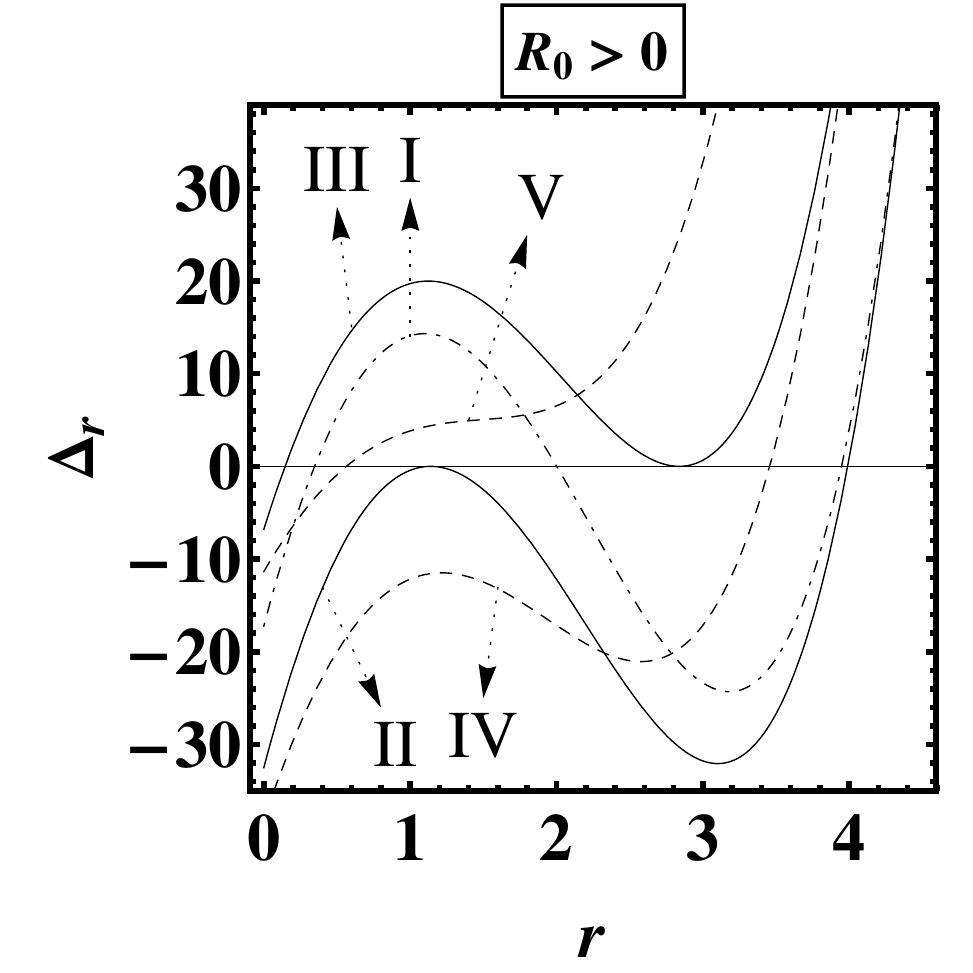}\,\,\,\,\,\,
		\caption{\footnotesize{
		Graphics showing positions of horizons as solutions of the equation $\Delta_r=0$. On the left panel ($R_0<0$) we show the cases $h>0$ ({\bf I}, BH with well-defined horizons, dashed with dots), $h=0$ ({\bf II}, {\it extremal} BH, continuous line) and $h<0$ ({\bf III}, {\it naked singularity}, dashed). On the right panel ($R_0>0$) we represent the cases $h<0$ ({\bf I}, BH with well-defined horizons, dashed with dots), $h=0$ ({\bf II}, $extremal$ BH and {\bf III}, $extremal$ $marginal$ BH, continuous line), and $h>0$ ({\bf IV}, {\it naked singularity} and {\bf V}, {\it naked marginal singularity}, dashed).}}
\end{figure}

To study the different thermodynamical properties of Kerr-Newman BH in $f(R)$ theories, we consider the Euclidean $f(R)$-Maxwell action in order to obtain the remaining thermodynamical quantities:
\begin{eqnarray}
\Delta S_E=\frac{1}{16 \pi}\int \text{d}^{4}x\sqrt{\mid g\mid}\,\left(R_0+f(R_0)-F_{\mu\nu}F^{\mu\nu}\right)\,,\nonumber
\end{eqnarray}

We can distinguish between two kinds of BH on this subject depending on the values of the $a$, $q$ and $M$ parameters and scalar curvature $R_0$: $i)$ {\it fast} BH, without phase transitions and always positive heat capacity. $ii)$ {\it slow} BH, which presents two phase transitions for two determined values of the horizon radius.

The temperature, entropy, heat capacity and the free energy of these BH happen to be:
\begin{eqnarray}
T_E= \frac{\displaystyle r_{ext}\left[1-\frac{R_0\,a^2}{12}-\frac{R_0\,r_{ext}^2}{4}-\frac{\left(a^2+Q^2\right)}{r_{ext}^2}\right]}{\displaystyle 4\pi\left(r_{ext}^2+a^2\right)}\,,\,\,\,\,\,
S=(1+f'(R_0))\,\frac{ {\cal A}_H }{4}\,,\,\,\,\,\,C=T\left.\frac{\partial S}{\partial T}\right|_{R_0,a,Q}\nonumber
\end{eqnarray}

\begin{eqnarray}
F=(1+f'(R_0))\frac{\displaystyle \left[36\,Q^2+12\,r_{ext}^2+r_{ext}^4\,R_0+a^2\,(36-r_{ext}^2\,R_0)\right]}{\displaystyle  24\,r_{ext}\,\Xi}\,\nonumber
\end{eqnarray}

with $ {\cal A}_H$ the area of the horizon of the BH. The multiplicative factor $1+f'(R_0)$ has to be positive in order to assure a positive mass and entropy for this kind of BH.

\begin{figure}[tp]
	\centering
		\includegraphics[width=0.400\textwidth]{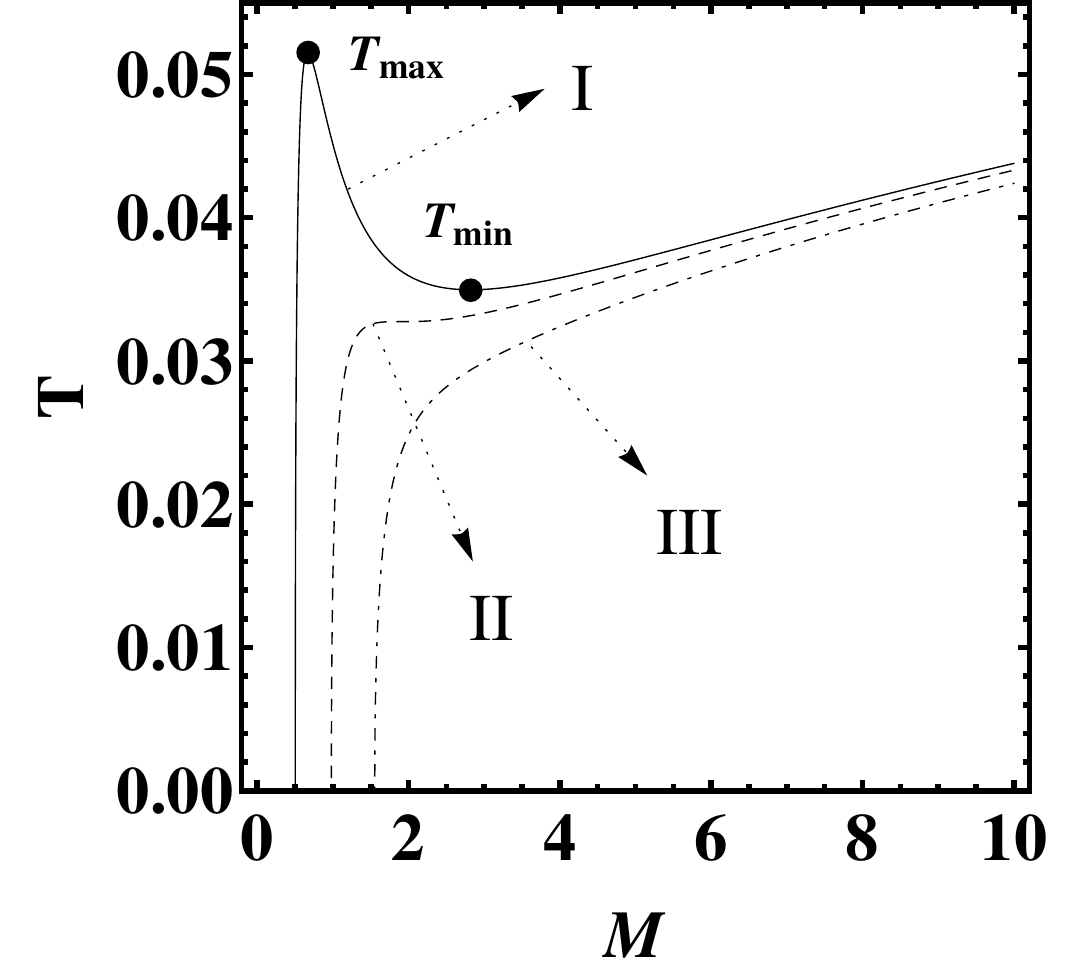}\,\,\,\,\,\,\,\,\,\,\,\,\,\,\,\,\,\,\,\,\,
		\includegraphics[width=0.43\textwidth]{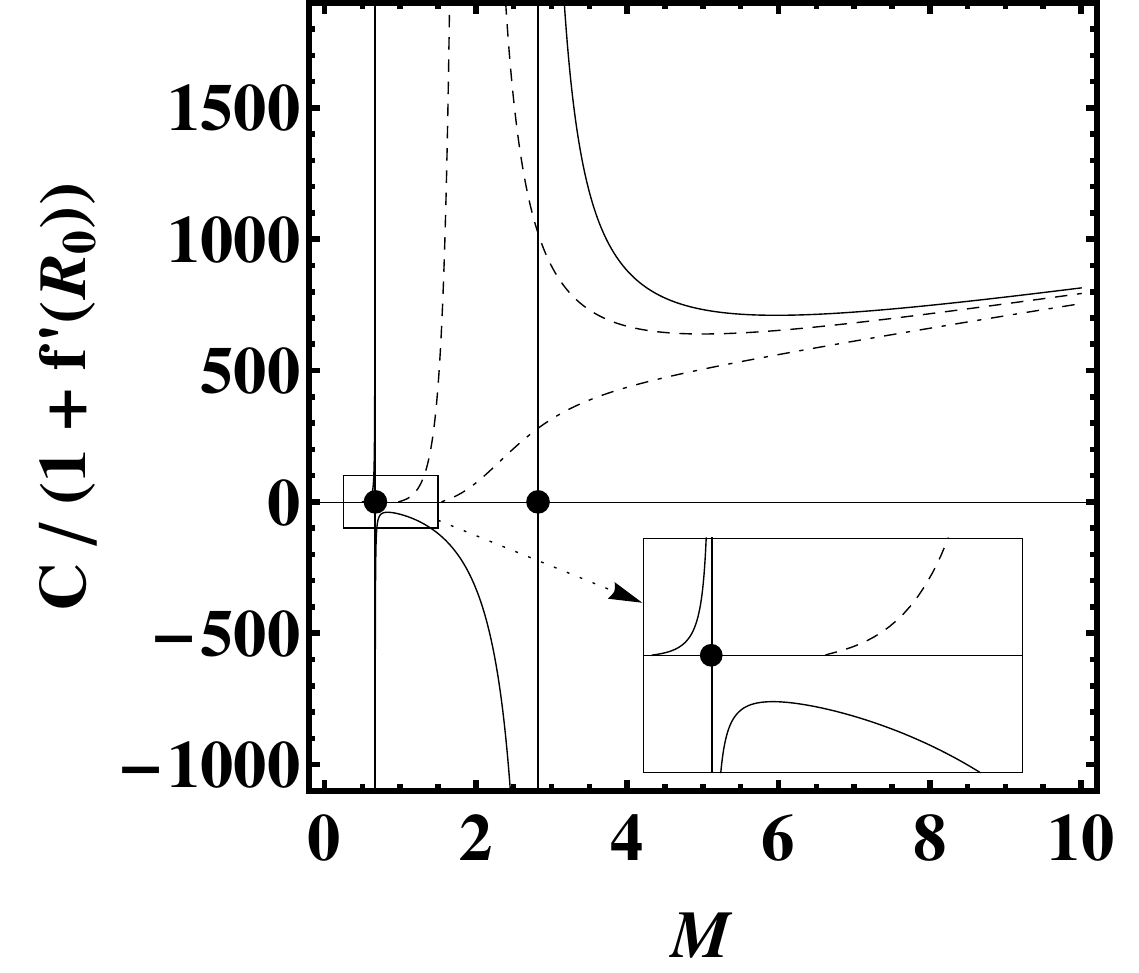}\,\,
\end{figure}

\begin{figure}[tp]
		 \includegraphics[width=0.41\textwidth]{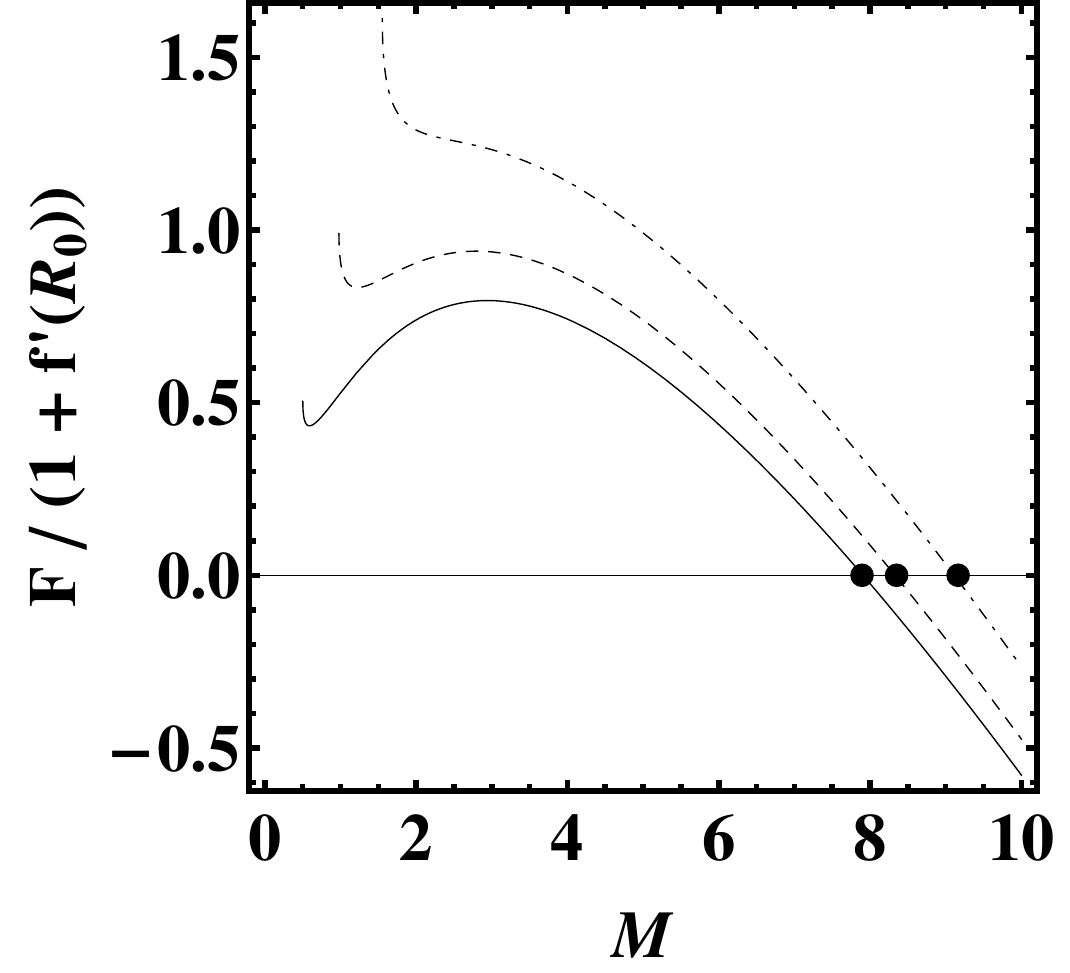}\,\,\,\,\,\,\,\,\,\,\,\,\,\,\,\,\,\,\,\,\,\,\,\,\,\,\,\,
		\includegraphics[width=0.3712\textwidth]{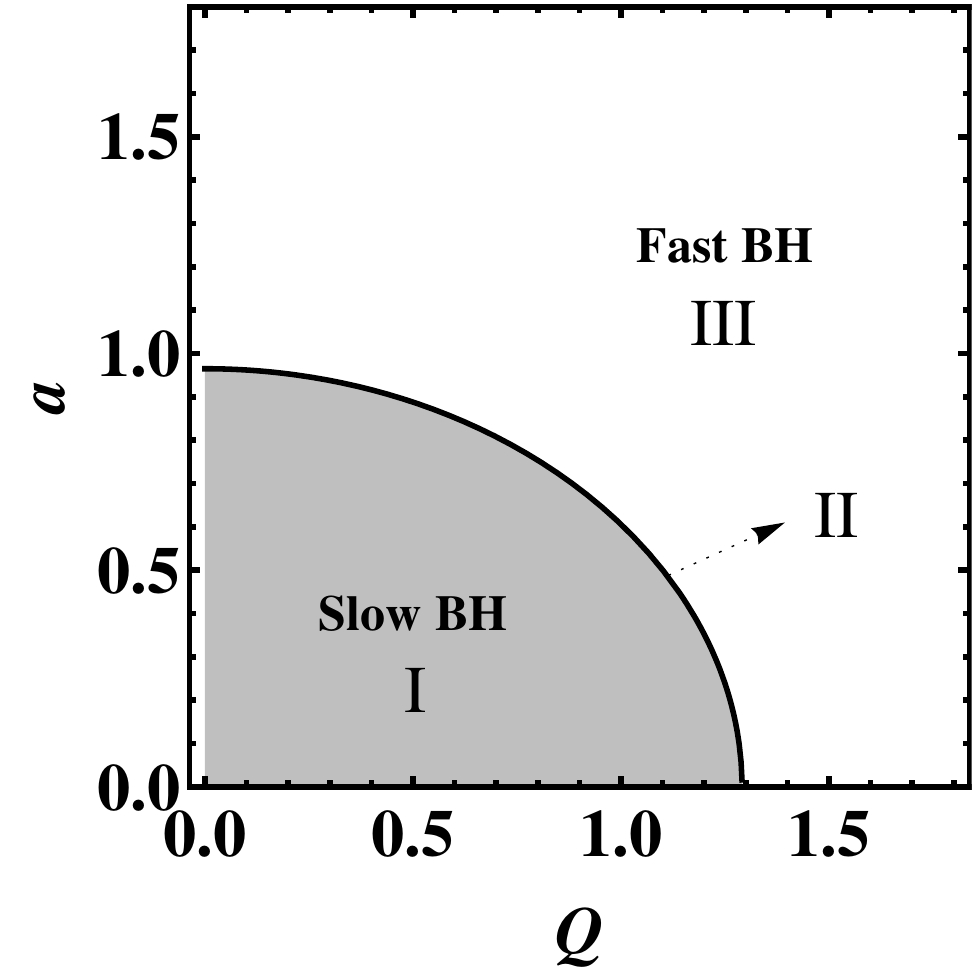}\,\,\,\,\,\,
		\caption{\footnotesize{For $R_0=-0.2$, we graphically display temperature (up to the left), heat capacity (up to the right), and the free energy (down to the left) of a BH as functions of the mass parameter $M$ for the cases: {\bf I)} $a=0.5$ and $Q=0$ : ``slow'' BH that shows a local maximum temperature $T_{max}$ and a local minimum temperature $T_{min}$ at the points where the heat capacity diverges, taking the latter negative values between $T_{max}$ y $T_{min}$. {\bf II)} $a\approx0.965$ and $Q=0$ : limit case where $T_{max}$ and $T_{min}$ merge, hence resulting on an inflection point in the temperature and an always positive heat capacity. {\bf III)} $a=1.5$ and $Q=0$ : ``fast'' BH with both temperature and heat capacity monotone growing (always positive too). It can be seen that all the configurations acquire a value $F<0$ from a certain value of $M$ onward, given by $r_{ext}^{\,\text{\it limit}}$. The values of $M^{min}$, with $T=0$ and $C=0$, correspond to an {\it extremal} BH. Down to the right, for $R_0=-0.2$ aswell, we display the regions in which BH behave as ``slow'' or ``fast'' BH.}}
\end{figure}

The existence of Kerr-Newman BH, as well as the thermodynamical quantities that define BH stability, depend upon the chosen $f(R)$ model range of parameters.

\begin{theacknowledgments}
This work has been supported by MICINN (Spain) project numbers FIS 2008-01323, FIS2011-23000, FPA 2008-00592, FPA2011-27853-01, Consolider-Ingenio MULTIDARK CSD2009-00064 (Spain) and URC (South Africa).
\end{theacknowledgments}





\bibliographystyle{aipproc}   




\end{document}